\documentclass[aps,prl,twocolumn,showpacs,nofootinbib,groupedaddress,superscriptaddress]{revtex4-1}

\usepackage{amssymb}
\usepackage{times}

\newcommand{\nc}{\newcommand}
\nc{\ox}{\otimes}

\newcommand{\1}{\openone}
\newcommand{\trace}{{\rm Tr}}

\newcommand{\set}[1]{{\left\{#1\right\}}}    % braces for set notation

\newcommand{\complex}{{\mathbb C}}

\usepackage{graphicx}% Include figure files
\usepackage{bm}% bold math
\usepackage{amsmath}
\usepackage{amsthm}

\newtheorem{theorem}{Theorem}

\newtheorem{definition}{Definition}

\newcommand{\Z}{the activation protocol}

\newcommand{\ket}[1]{|#1\rangle}
\newcommand{\bra}[1]{\langle#1|}
\newcommand{\proj}[1]{\ket{#1}\bra{#1}}
\newcommand{\braket}[2]{\langle #1 | #2 \rangle}
\newcommand{\wek}[1]{{\boldsymbol{#1}}}

\newcommand{\ot}[0]{\otimes}
\newcommand{\beq}{\begin{equation}}
\newcommand{\eeq}{\end{equation}}

\newcommand{\Tr}{{\rm Tr}}

\newcommand{\cB}{\mathcal{B}}

\newcommand{\cN}{\mathcal{N}}

\newenvironment{beweis}[1][{\hspace{-\blank}}]{{\noindent\emph{Proof~{#1}.\ }}}{\hfill\qed\vskip 0.5\baselineskip}
\nc{\CC}{{{\mathbb C}}}
\nc{\EE}{{{\mathbb E}}}

\begin{document}

\title{All nonclassical correlations can be activated into distillable entanglement}
%Activation of nonclassical correlations}
%Equivalence between multipartite nonclassical correlations and distillable entanglement between system and environment
%Activating nonclassicality: Equivalence between quantumness of correlations and distillable entanglement between system and environment.}
%Activation of nonclassical correlations: entanglement potential of the relative entropy of quantumness}

\author{Marco Piani}
\affiliation{Institute for Quantum Computing and Department of Physics and Astronomy, University of Waterloo, Waterloo
N2L 3G1, Canada}
\author{Sevag Gharibian}
\affiliation{Institute for Quantum Computing and School of Computer Science, University of Waterloo, Waterloo
N2L 3G1, Canada}
\author{Gerardo Adesso}
\affiliation{School of Mathematical Sciences, University of Nottingham, University Park, Nottingham NG7 2RD, United Kingdom}
\author{John Calsamiglia}
\affiliation{F\'{i}sica Te\`{o}rica: Informaci\'{o} i Fen\`{o}mens Qu\`{a}ntics, Universitat Aut\`{o}noma de Barcelona, 08193 Bellaterra, Spain}
\author{Pawe{\l} Horodecki}
\affiliation{Faculty of Applied Physics and Mathematics, Technical University of Gda\'nsk, 80-952 Gda\'nsk, Poland}
\affiliation{National Quantum Information Centre of Gdansk, 81-824, Sopot, Poland}
\author{Andreas Winter}
\affiliation{Department of Mathematics, University of Bristol, Bristol BS8 1TW, United Kingdom \\ and Centre for Quantum Technologies, National University of Singapore, Singapore 117542}

\begin{abstract}
%The correlations of multipartite quantum states have nonclassical features that go beyond entanglement.
We devise a protocol in which general nonclassical multipartite correlations produce a physically relevant effect, leading to the creation of bipartite entanglement. In particular, we show that the relative entropy of quantumness, which measures all nonclassical correlations
 among subsystems of a quantum system, is equivalent to and can be operationally interpreted as the minimum distillable entanglement generated between the system and local ancillae in our protocol.
%All and only the nonclassically correlated states  possess an entanglement potential and are thus useful for quantum information processing.
We emphasize the key role of state mixedness in maximizing nonclassicality: Mixed entangled states can be arbitrarily more nonclassical than separable and pure entangled states.
\end{abstract}


\date{May 10, 2011}

\maketitle
%\section{Introduction}
The study of quantum correlations has traditionally  focused on entanglement~\cite{RevModPhys.81.865}. It is generally believed that entanglement is a necessary resource for quantum computers to outperform their classical counterparts. Indeed, it has been shown that for the setting of \emph{pure}-state computation, the amount of entanglement present must grow with the system size for an exponential speed-up to occur~\cite{josza}. In the context of \emph{mixed}-state quantum information processing, however, there are  computational and communication feats which are seemingly impossible to achieve with a classical computer, and yet can be attained with a quantum computer using little or no entanglement (e.g.~\cite{PhysRevLett.81.5672,DHLST04}). For example, the Deterministic Quantum Computation with one Qubit (DQC1) model is believed to estimate the trace of a unitary matrix exponentially faster than any classical algorithm, yet with vanishing entanglement during the computation~\cite{PhysRevLett.100.050502}. A second example is the ability for certain bipartite quantum systems to contain a large amount of ``locked'' classical correlations, which can then be ``unlocked'' with a disproportionately small amount of classical communication~\cite{DHLST04}. This task is impossible classically,  yet the quantum states involved are \emph{separable}, that is, unentangled. This raises the crucial question about which, if not entanglement,  is the fundamental resource enabling such feats.

%Despite the afact that the study of quantum correlations has traditionally focused on entanglement~\cite{RevModPhys.81.865}, there is much interest recently in nonclassical correlations that are more general than entanglement. Such an interest has been in good part triggered by the potential connection between nonclassicality of correlations and the computational speed-up in the model known as \emph{deterministic quantum computation with one quantum bit} (DQC1)~\cite{PhysRevLett.81.5672}.
%The DQC1 model, or one-clean-qubit model, is a model of quantum computation in which all but one qubit start in the maximally mixed state.  It is worth remarking that while (a growing, along the computation) entanglement is needed for any speed-up in pure-state quantum computation~\cite{josza}, little is known about the role of entanglement in mixed-state quantum computation. DQC1 computers are believed to be weaker than standard---e.g., circuit-model---quantum computers, but still able to solve efficiently some  problems that appear to be classically intractable. What is intriguing is that the power of the model does not rely on the  generation of a large amount of entanglement. Indeed, the latter, depending on the qubits considered, is found to be little or even vanishing along the computation. It was thus suggested that the presence of nonclassical correlations other than entanglement, in particular between the clean qubit and the other qubits, is at the origin of the speed-up~\cite{PhysRevLett.100.050502}.

One plausible explanation is associated with the presence in (generic~\cite{ferraro}) quantum states of correlations which have nonclassical signatures that go \emph{beyond} entanglement.
Indeed, much attention has recently been devoted to understanding and quantifying such correlations for this very reason
\cite{PhysRevLett.88.017901,PhysRevA.71.062307,PhysRevA.72.032317,PhysRevLett.104.080501,groismanquantumness,PhysRevA.77.052101,Bravyi2003,pianietal2008nolocalbrodcast,pianietal2009broadcastcopies,ferraro,ADA}.
In particular, the separable quantum states of the systems involved in DQC1 and the locking protocol have been shown to possess non-zero amounts of such correlations \cite{PhysRevLett.100.050502,DG09}, as measured by the {\it quantum discord}
\cite{PhysRevLett.88.017901}. The latter strives to capture nonclassical correlations beyond entanglement and has recently received operational interpretations in terms of the quantum state merging protocol~\cite{operdiscord}, but is unfortunately not a \emph{faithful} measure \cite{notefaithful}.  A more accurate quantification of nonclassical correlations is provided by the so-called {\em relative entropy of quantumness} (REQ) \cite{PhysRevA.71.062307,Bravyi2003,groismanquantumness,PhysRevA.77.052101,PhysRevLett.104.080501}, defined as the minimum distance, in terms of relative entropy, between a multipartite quantum state and the closest strictly classically correlated state (see Definition~\ref{def:classical}). Such a measure is faithful \cite{groismanquantumness}, symmetric under permutation of the subsystems, and enables a unified approach to the quantification of classical, separable and entangled correlations \cite{PhysRevLett.104.080501}.

 More generally, the role of nonclassical correlations in quantum information tasks remains unclear.
 While all entangled states are known to be useful for  information processing \cite{Masanes}, the fundamental question of whether the same holds for all nonclassically correlated (separable) states stays open.
This raises the question:
Is there a setting in which general nonclassical correlations produce a physically relevant effect that distinguishes them from purely classical ones?
%\vspace{2mm}

%In this Letter, we answer the question in the affirmative by demonstrating a protocol which in some sense \emph{activates} the nonclassicality present in any multipartite quantum system, leading to the creation of entanglement. As an application, the REQ measuring nonclassical correlations in the state of a system is proven to be equivalent to the minimum distillable entanglement generated between the  system and local ancillae in our protocol. This result renders the REQ \emph{both} an operational and faithful nonclassicality measure.
%%The framework we introduce provides a natural and general link between entanglement monotones (quantifying the activated system-ancilla entanglement at the output of the protocol) and nonclassicality measures (quantifying the input intra-system correlations).
%In general, all and only the quantumly correlated states are shown to possess an entanglement potential that makes them readily useful for better-than-classical information processing. We further prove that nonclassical correlations are limited for separable states and for pure entangled states in any dimension, and are instead maximized on typical {\it mixed}, entangled states.

In this Letter, we answer the question in the affirmative by demonstrating a protocol which in some sense \emph{activates} the nonclassicality present in any multipartite quantum system, leading to the creation of entanglement. We then show that the REQ of any system state input to our protocol is precisely the minimum distillable entanglement generated between the system and local ancillae via the protocol. This result  renders the REQ \emph{both} an operational \emph{and} faithful nonclassicality measure. According to our framework, all and only the quantumly correlated states are shown to possess an entanglement potential that makes them readily useful for better-than-classical information processing. Finally, we prove limits on nonclassical correlations for separable and pure entangled states in any dimension, while, perhaps surprisingly, these bounds can be exceeded by {\it mixed} entangled states.

Our results apply to general multipartite states, adopting the following definition of classicality  \cite{pianietal2008nolocalbrodcast}.
% in a multipartite quantum system.
%\vspace{-1.7mm}
\begin{definition}[Strictly Classically Correlated Quantum State]
\label{def:classical}
Given a set of n $d$-dimensional qudit systems, let ${\cB}_i$ denote an orthonormal basis in $\complex^d$ for the $i$th system consisting of vectors $\ket{\cB_i(k)}$ for $0\leq k\leq d-1$, and let ${\cB}$ denote an orthonormal basis $\{\ket{\cB(\wek{k})}=\ket{\cB_1(k_1)}\ket{\cB_2(k_2)}\cdots \ket{\cB_n(k_n)}\}$ for the entire space $(\complex^{d})^{\otimes n}$ formed by taking tensor products of all elements in bases $\set{{\cB}_i}_{i=1}^n$.
%(e.g. $\ket{B_1(1)}\otimes \ket{B_2(1)}\otimes\cdots\otimes \ket{B_n(1)}\in C$).
Then, an $n$-qudit state $\rho$ is \emph{strictly classically correlated}---or  simply \emph{classical}---if there exists such a basis $\cB$ with respect to which $\rho$ is diagonal. Such states correspond to the embedding of a multipartite classical probability distribution into the quantum formalism.
\end{definition}

%\noindent{\bf Mapping nonclassical correlations to entanglement.}
{\emph{Activation protocol.}}---
We now describe our protocol for activation of nonclassical correlations. The scheme is somewhat inspired by the quantum optics setup of~\cite{john}, where  one attempts to quantify nonclassicality of a single field mode (defined there as the state deviation from a mixture of coherent states) by reducing the problem to quantifying the two-mode \emph{entanglement} that can be generated from
the field using linear optics, auxiliary classical (coherent) states, and ideal photodetectors.
Similarly, we may expect that  mapping the (still not-well-understood)  nonclassicality of multipartite \emph{correlations} into ``more familiar'' bipartite entanglement allows one to employ tools from entanglement theory \cite{RevModPhys.81.865} to interpret and quantify general nonclassical correlations.

\begin{figure}[t]
\includegraphics[width=8.5cm]{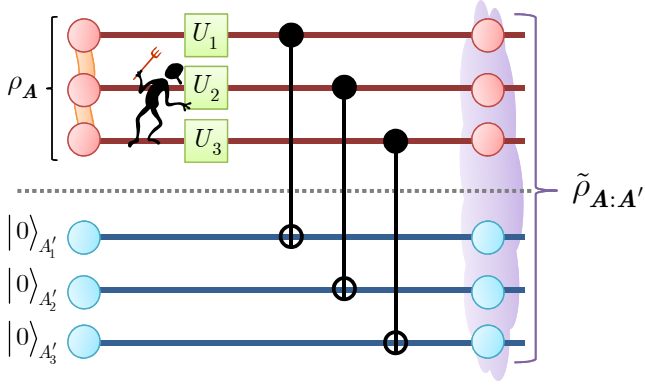}
\caption{(Color online) Activation protocol for $n=3$.}\label{figact}
\end{figure}

Our activation protocol can be thought of as a game between an adversary and $n$ players, where the $n$ players together aim to generate an entangled state between a system $\wek{A}$ they control and an ancillary system $\wek{A'}$, and the adversary's goal is to thwart their efforts by locally rotating each subsystem of $\wek{A}$ before system and ancilla undergo a pre-defined interaction.
More precisely, the  protocol proceeds as follows (see Fig.~\ref{figact}).
We consider $n$ players ${\cal P}_i$, each controlling a system-ancilla pair of qudits $(A_i, A'_i)$.
We indicate by $\wek{A}$ the joint register  $A_1,\ldots,A_n$ (``system''), and by $\wek{A'}$ the joint register $A'_1,\ldots,A'_n$ (``ancilla''). The initial state of the total $2n$ qudits is a tensor product  $\rho_{\wek{A}:\wek{A'}} = \rho_{\wek{A}} \otimes \ket{0}\bra{0}^{\otimes n}_{\wek{A'}}$.
For a given  $\rho_\wek{A}$, an adversary is first allowed to apply a local unitary $U_i$ of his choice to each $A_i$.
With the adversary's turn complete, each player ${\cal P}_i$ now  lets their subsystem $A_i$ (control qudit) interact with the corresponding ancillary party $A'_i$ (target qudit) via a \textsc{cnot} gate $C_{A_i:A'_i}$,
whose action  on the computational basis states $\ket{j}\ket{j^\prime}$ of $\complex^d \otimes \complex^d$ is defined as
$C\ket{j}\ket{j^\prime}=\ket{j}\ket{j^\prime\oplus j}$,
with $\oplus$ denoting addition modulo $d$. The final state of system plus ancilla is
\begin{equation}\label{eq:finalstate}
\tilde{\rho}_{\wek{A}:\wek{A'}} = V (\rho_{\wek{A}} \otimes \ket{0}\bra{0}^{\otimes n}_{\wek{A'}}) V^\dagger\,,
\end{equation}
with $V = C_{\wek{A}:\wek{A'}} \cdot (U_{\wek{A}} \otimes \openone_{\wek{A'}})$, $U_{\wek{A}} = \otimes_{i=1}^n U_i$, and $C_{\wek{A}:\wek{A'}} = \otimes_{i=1}^n C_{A_i:A'_i}$.
We ask: At the end of the protocol, have the $n$ players succeeded in generating bipartite entanglement across the split $\wek{A}:\wek{A'}$, and, if so, how much entanglement was created? It is natural to expect that the answer will depend on the initial state $\rho_{\wek {A}}$ of the $n$-qudit system. From a physical perspective, our aim is to understand precisely how the nature and amount of correlations between the parts $A_i$ of the system $\wek{A}$ affects the entanglement that can be created with an ancilla $\wek{A'}$ via the paradigmatic entangling operation --- the \textsc{cnot}; we consider the worst case scenario with respect to the choice of the control bases. We then find the following.

\begin{theorem}
\label{thm:quantumnessiff}
The (initial) state $\rho_{\wek{A}}$ of an $n$-qudit system is strictly classically correlated if and only if there exists some adversarial choice of local unitaries $U_{\wek{A}}$ such that the (final) state $\tilde{\rho}_{\wek{A}:\wek{A'}}$ output by the activation protocol is separable across the system-ancilla split.
\end{theorem}
In other words, the system \emph{always} becomes (for any choice of $U_{\wek{A}}$) entangled with the ancilla as a result of the activation protocol, if and only if the input state of the system is nonclassically correlated. This establishes a qualitative \emph{equivalence} between multipartite nonclassical correlations among components of a quantum system, and bipartite entanglement between the system and an ancilla, and settles the issue of the usefulness of nonclassical correlations in (even separable) quantum states for quantum primitives: Any kind of \emph{multipartite} nonclassicality initially present in $\wek{A}$ is a resource for information processing that can always be activated, or mapped into \emph{bipartite} entanglement across the $\wek{A}:\wek{A^\prime}$ split.  While a direct proof of this result is quite straightforward (see Appendix A \cite{epaps}), in the following we show a more powerful result that promotes the equivalence between
%input
nonclassicality and
%output
entanglement to a quantitative relationship.

%To prove this, one can use arguments very similar to those of \cite{john}. We do not report explicitly such a proof because in the following we will actually be able to quantify the non-vanishing minimum entanglement generated if the state is not classical.
%On the other hand, it is easy to see that if the state is striclty classically correlated, we can choose the unitaries to make the factorized basis $\wek{\cB}$ in which the state is diagonal coincide with the control bases of the CNOT gates, so that entanglement is not created.

{\emph{Quantifying nonclassicality.}}--- Having run \Z, we now proceed to the next logical step: Namely, we wish to \emph{quantify} the entanglement generated in the $\wek{A}:\wek{A^\prime}$ split whenever $\wek{A}$ is initially in a nonclassically correlated state. The present framework is general enough to allow us to uncover a full zoology of nonclassicality measures, as each choice of a different entanglement monotone~\cite{reviewplenio} we adopt (at the output) leads in principle to a unique nonclassicality measure (for the input state), the  association stemming exactly from the activation protocol. More precisely, let $E$ denote some entanglement measure of choice and $\tilde{\rho}_{\wek{A}:\wek{A'}}$ the final system-ancilla state  as in Eq.~(\ref{eq:finalstate}), and define by \begin{equation}\label{eq:QE}Q_E({\rho}_{{\wek{A}}}):=\min_{U_{{\wek{A}}}}E_{\wek{A}:\wek{A'}}(\tilde{\rho}_{\wek{A}:\wek{A'}})\,
 \end{equation}
 the minimum entanglement generated across the $\wek{A}:\wek{A^\prime}$ split over all choices of adversarial local unitaries $U_\wek{A}$. We call $Q_E({\rho}_{\wek{A}})$ the \emph{minimum entanglement potential} of ${\rho}_{\wek{A}}$ with respect to $E$. As a consequence of Theorem~\ref{thm:quantumnessiff}, $Q_E$ is a  measure of nonclassical correlations in the multipartite state $\rho_{\wek{A}}$, for every entanglement monotone $E$.

In fact, the condition  $Q_E({\rho}_{\wek{A}})=0$ perfectly characterizes the set of classically correlated states $\rho_\wek{A}$ if $E$ is a \emph{faithful} entanglement measure (i.e. if $E$ vanishes only for separable states). However, even certain non-faithful entanglement measures can be plugged in to obtain a faithful measure of nonclassical correlations~\cite{notefaithful}.
%(analogously, a faithful nonclassicality measure vanishes if and only if a state is strictly classically correlated).
The reason is that the output state $\tilde{\rho}_{\wek{A}:\wek{A'}}$ has the so-called \emph{maximally correlated} form~\cite{rains2001} between $\wek{A}$ and $\wek{A'}$; namely, ${\tilde{\rho}}_{\wek{A}:\wek{A'}}=\sum_{\wek{k}\wek{l}}\rho_{\wek{k}\wek{l}}^\wek{\cB}\ket{\wek{k}}\bra{\wek{l}}_{\wek{A}}\otimes \ket{\wek{k}}\bra{\wek{l}}_{\wek{A'}}$
with $\rho_{\wek{k}\wek{l}}^\wek{\cB}=\bra{\wek{\cB}(\wek{k})}\rho_\wek{A} \ket{\wek{\cB}(\wek{l})}$,
$\ket{{\cB}(\wek{k})}_{\wek{A}}=U_\wek{A}^\dagger \ket{\wek{k}}$ and $\ket{\wek{k}}=\ket{k_1}\ket{k_2}\cdots\ket{k_n}$.
%We hence can obtain an \emph{operational} measure of nonclassical correlations, which unlike the quantum discord, is also faithful.
In particular, let us consider the non-faithful (as it vanishes on so-called bound entangled states) but physically motivated distillable entanglement $E_{\textup{D}}$~\cite{reviewplenio} as a bipartite entanglement monotone. We find that the $\wek{A}:\wek{A'}$ distillable entanglement of $\tilde{\rho}_{\wek{A}:\wek{A'}}$ is equal to
$E_{\textup{D}}(\tilde{\rho}_{\wek{A}:\wek{A'}})=S(\tilde{\rho}_{\wek{A}})-S(\tilde{\rho}_{\wek{A}:\wek{A'}})=S({\rho}_{\wek{A}}^\wek{\cB})-S(\rho_{\wek{A}})
						%&=H(\{\rho^u_{i,i}\})-S(\rho_{\wek{A}})
$,
where $S(\sigma)=-\Tr( \sigma \log_2 \sigma)$ is the von Neumann entropy of a state $\sigma$. In the first equality we used the results of~\cite{Hiroshima2004} about distillable entanglement for maximally correlated states---for which it happens to coincide with the relative entropy of entanglement~\cite{PhysRevLett.78.2275}. The second equality is justified by the fact that $\rho_{\wek{A}}^{\wek{{\cB}}}$ is the state resulting from local projective measurements in the local bases $\wek{\cB}$ on $\rho_{\wek{A}}$ and is unitarily equivalent to $\tilde{\rho}_{\wek{A}}$, while $\tilde{\rho}_{\wek{A}:\wek{A'}}$ is obtained from $\rho_{\wek{A}}$ via the activation protocol isometry, Eq.~(\ref{eq:finalstate}). Thus, the minimum distillable entanglement potential $Q_{E_D}({\rho}_{\wek{A}})$ takes on the form
%\beq
%\label{eq:localignorance}
$Q_{E_{\textup{D}}}(\rho_{\wek{A}})=\min_{{\cB}}\left(S(\rho_{\wek{A}}^{\wek{\cB}})-S(\rho_{\wek{A}})\right)$,
%\eeq
where the minimization is over the choice of the bases $\wek{\cB}$. As proven in~\cite{PhysRevLett.104.080501}, this is  an equivalent expression for the REQ,
\beq
\label{eq:req}
Q(\rho_{\wek{A}})=\min_{\textrm{classical}~\sigma_\wek{A}}S(\rho_\wek{A}\|\sigma_\wek{A}),
\eeq
where the relative entropy is defined as $S(\rho\|\sigma)=\trace(\rho \log_2 \rho - \rho \log_2 \sigma)$ and the minimization is over all strictly classically correlated states $\sigma_\wek{A}$.
We have thus proven that the REQ quantifying general nonclassical correlations between the $n$ subsystems $A_i$ of $\wek{A}$
is exactly equal to the minimum bipartite distillable entanglement potential---or, equivalently, to the minimum relative entropy of entanglement potential---generated between the system $\wek{A}$ and the ancillary register  $\wek{A'}$.
% by means of our activation protocol.

This finding immediately provides a clearcut \emph{operational interpretation} for the REQ, a quantity whose original definition was purely geometric [Eq.~(\ref{eq:req})], which then emerges as a mathematically sound  and physically motivated measure of nonclassical correlations for arbitrary quantum states, quantifying equivalently the resource power of such correlations for (distillable) entanglement generation.
Incidentally, since the REQ is faithful \cite{groismanquantumness}, this yields a proof of Theorem~\ref{thm:quantumnessiff}.

Other  nonclassicality measures can be induced by different entanglement monotones. Choosing e.g.~the ``negativity'' $\cN$~\cite{VW02} as an entanglement measure, one obtains  $Q_\cN(\rho_\wek{A})=(\min_{\wek{\cB}} \sum_{\wek{i}\neq \wek{j}}|\rho^{\wek{\cB}}_{\wek{i},\wek{j}}|)/2$ as a quantifier of nonclassical correlations (see Appendix B~\cite{epaps} for details), directly related to the off-diagonal coherences of the density matrix of the system, minimized over all local bases.

\emph{Nonclassicality versus mixedness and entanglement.}---
Equipped with a faithful and operational measure of nonclassical correlations, the REQ  $Q\equiv Q_{E_{\textup{D}}}$, we can
%leave aside the ancillae for the remainder of this Letter and focus in detail on the composite qudit system $\wek{A}$ in order to
investigate the interplay between nonclassicality, entanglement and mixedness of general states  $\rho_{\wek{A}}$. For the sake of simplicity, from now on we restrict to the bipartite case $A_1=A$, $A_2=B$.  We begin with a few simple but general observations following from the definition of $Q$.

For \emph{pure} states $\rho_{AB} = \proj{\psi}$, the quantumness $Q$ reduces to the von Neumann entropy of entanglement $S(\rho_A)=S(\rho_B)$ \cite{Bravyi2003}, and is thus at most equal to  $\log_2 d$. On the other hand, for arbitrary mixed $\rho_{AB}$, we have that $Q(\rho_{AB})$ is at most $2\log_2 d$, since from Eq.~(\ref{eq:req}) one has $Q(\rho_{AB}) \le S(\rho_{AB}\|\rho_A \otimes \rho_B) = S(\rho_A)+S(\rho_B)-S(\rho_{AB}) \equiv I(\rho_{AB})$, where $I$ denotes the mutual information, a measure of \emph{total} correlations. From this and the results of~\cite{nielsenkempe}, one realizes that for a \emph{separable} state a bound $Q(\rho_{AB}^{\textup{sep}}) \leq\log_2 d$ holds.
In Appendix C \cite{epaps}, we prove in fact that this inequality is always sharp for separable states, i.e., the bound $\log_2 d$ cannot be exactly saturated for separable nonclassical states, while it is instead trivially reached by pure maximally entangled states $\ket{\psi} = d^{-1/2} \sum_{j=0}^{d-1} \ket{j}\ket{j}$. Almost all separable states thus possess nonclassical correlations \cite{ferraro}, but not to a maximal extent (as already observed in the particular cases of  two-qubit \cite{alqasimi} and two-mode Gaussian states \cite{ADA}). However, with increasing $d \rightarrow \infty$ we find quite surprisingly that the upper bound on the REQ of separable states becomes asymptotically tight, in the sense that separable states exist such that $Q(\rho^{\textup{sep}}_{AB}) / \log_2 d \rightarrow 1$. Even more intriguingly, we can show that the upper bound on general mixed bipartite states $\rho_{AB}$ is also asymptotically tight, in the sense that families of mixed states exist such that in the limit $d \rightarrow \infty$, their quantumness converges to the maximum, $Q(\rho_{AB}) / \log_2 d\rightarrow 2$.
More precisely, in Appendix D
\cite{epaps} we prove the following two results using techniques from Refs.~\cite{HLSW:rand,HLW:aspects}. Let $m = \lceil (\log_2 d)^4 \rceil$.
\begin{theorem}
  \label{prop:sep}
  Define the following random separable state:
   $ \sigma_{AB} = \frac{1}{dm} \sum_{{i=1,\ldots,d}\atop{j=1,\ldots,m}}
                                    \proj{i}_{A} \ox \left(U_j\proj{i}U_j^\dagger\right)_{B}$,
  with unitaries $U_j$ drawn independently from the Haar measure.
  Then,
    $S(\sigma_{AB}) \leq \log_2 d + \log_2 m$,
  while on the other hand, for $d$ sufficiently large and with high probability,
     $S\bigl(\sigma^{\wek{\cB}}_{AB} \bigr) \geq 2\log_2 d - \textup{const.}$, for all $\wek{\cB}$.
  Hence, $Q(\sigma_{AB}) \geq \log_2 d - O(\log_2 \log_2 d)$.
\end{theorem}

\begin{theorem}
  \label{prop:low-rank}
  Define the following random state: For $C$ a system of dimension $m$, let
    $\rho_{AB} = \Tr_C \proj{\psi}_{ABC}$,
  where $\ket{\psi} \in \mathbb{C}^d \ox \mathbb{C}^d \ox \mathbb{C}^m$ is uniformly distributed
  (with probability induced by the Haar measure).
  Then, $S(\rho) \leq \log_2 m$,
  while on the other hand, for $d$ sufficiently large and with high probability, $S\bigl( \rho^\wek{\cB} \bigr) \geq 2\log_2 d - \textup{const.}$, for all $\wek{\cB}$.
  Hence, $Q(\rho_{AB}) \geq 2\log_2 d - O(\log_2 \log_2 d)$.
\end{theorem}
%\begin{proof}
%See the Appendix.
%\end{proof}

These results show that, first, there are separable states that asymptotically (in $d$) are as nonclassical as the most nonclassical pure state (which is the maximally entangled state); second, mixed entangled states can be twice as  nonclassical as pure entangled states. \emph{Both} entanglement \emph{and} mixedness are  required to ``break the barrier'' of $\log_2 d$, thus showing that entanglement by itself is not the strongest form of nonclassicality.

\emph{Conclusions.}--- The study of general nonclassical correlations is currently a burgeoning area, but in many ways such correlations are still not well-understood. Our activation protocol lends new insight into the nature of these correlations by furnishing them, in full generality, with a new operational meaning in terms of resources for \emph{entanglement generation}.
Furthermore,  we have reduced the problem of quantifying nonclassicality  to the more familiar setting of quantifying entanglement, for which a multitude of tools for analysis are already known (see e.g.~\cite{RevModPhys.81.865}). As an added bonus, we have obtained an alternative operational interpretation for the relative entropy of quantumness measure \cite{PhysRevA.71.062307,PhysRevLett.104.080501}. Finally, with respect to the latter, we have demonstrated that, remarkably, there exist \emph{mixed} entangled quantum states whose nonclassical correlations are \emph{stronger} than those of pure entangled states.
Further investigation on the nature and the structure of nonclassical correlations, following the programme laid by this Letter, may trigger novel developments in quantum technology  and shed light on foundational aspects of quantum theory.

 {\it Note added.---}After completion of this Letter, we became aware of some related results by Streltsov et al.~\cite{streltsov2010}, who showed that the quantumness of correlations (as measured e.g.~by the quantum discord) is also related to the minimum entanglement generated between system and apparatus in a partial measurement process. In light of those results, our findings can be understood also as dealing with the interplay between system-apparatus entanglement and nonclassicality of correlations when realizing local measurements.

We thank  F.~Brand\~ao, N.~Brunner, D.~Bru\ss,  H.~Kampermann, D.~Leung and A.~Streltsov for discussions. We acknowledge support by NSERC, QuantumWorks,
CIFAR, Ontario Centres of Excellence, the Spanish government (program FIS2008-01236/FIS), the Catalan government (program 2009SGR-0985), the United Kingdom EPSRC, the European Commission, the ERC, the Philip
Leverhulme Trust, the Royal Society, and the Integrated
Project QESSENCE. M. P. was supported by the Austrian
Science Fund (FWF) through the Lise Meitner program
while at the University of Innsbruck.

\cleardoublepage
\appendix
\begin{widetext}

{\noindent \large{Supplemental Material}}

\begin{center}
{\large \bf
 All non-classical correlations can be activated into distillable entanglement}

 \smallskip

Marco Piani, Sevag Gharibian, Gerardo Adesso, John Calsamiglia, Pawe{\l} Horodecki, and Andreas Winter

\smallskip

\end{center}

 \setcounter{page}{1}
  \pagenumbering{roman}

\section{Proof of Theorem \ref{thm:quantumnessiff}}
\label{app:Proof}

\begin{proof} The ``if'' part is trivial, as given a strictly classical correlated state one can choose $U_{\wek{A}}$ to make the $n$-orthogonal spectral basis of
\[
\rho_{\wek{A}}=\sum_{\wek{i}}p_{\wek{i}}\proj{\wek{\cB}(\wek{i})}
\]
coincide with the computational basis, so that
\[
\tilde{\rho}_{\wek{A}:\wek{A'}}=\sum_{\wek{i}}p_{\wek{i}}\proj{\wek{i}}_{\wek{A}}\otimes\proj{\wek{i}}_{\wek{A'}}.
\]
As regards the ``only if'' part, let us consider the separable decomposition
\[
\tilde{\rho}_{\wek{A}:\wek{A'}}=\sum_\alpha q_\alpha\proj{\psi^\alpha}_{\wek{A}}\ot \proj{\phi^\alpha}_{\wek{A'}}
\]
which exists by hypothesis for some choice of $U_{\wek{A}}$.
Since the transformation \eqref{eq:finalstate} is unitary and invertible, there must exist a pure ensemble $\{q_\alpha,\ket{\xi^\alpha}_{\wek{A}}\}$ such that $\rho_{\wek{A}}=\sum_\alpha{q_\alpha}\proj{\xi^\alpha}_{\wek{A}}$ and
\beq
\label{eq:purecond}
V\ket{\xi^\alpha}_{\wek{A}}\ket{0}_{\wek{A'}}=\ket{\psi^\alpha}_{\wek{A}}\ot \ket{\phi^\alpha}_{\wek{A'}}
\eeq
Let us expand $\ket{\xi^\alpha}_{\wek{A}}$ on the computational basis rotated by $U^\dagger_{\wek{A}}$:
\[
\ket{\xi^\alpha}_{\wek{A}}=\sum_{\wek{a}}c^\alpha_{\wek{a}_{\wek{A}}}U^\dagger_{\wek{A}}\ket{\wek{a}}_{\wek{A}}
\]
and compute the action of $V$:
\[
V\ket{\xi^\alpha}_{\wek{A}}\ket{0}_{\wek{A'}}=\sum_{\wek{a}}c^\alpha_{\wek{a}_{\wek{A}}}\ket{\wek{a}}_{\wek{A}}\ot\ket{\wek{a}}_{\wek{A'}}
\]
Imposing the factorization condition \eqref{eq:purecond} we find that it must be $c^\alpha_{\wek{a}_{\wek{A}}}=c^\alpha_f(\alpha)\delta_{\wek{a},f(\alpha)}$, $|c^\alpha_f(\alpha)|=1$ for some $f(\alpha)\in \mathbb{N}^n$. Therefore,
\[
\begin{split}
\rho_{\wek{A}}
&=\sum_\alpha{q_\alpha}\Big(\sum_{\wek{a}}c^\alpha_{\wek{a}_{\wek{A}}}U^\dagger_{\wek{A}}\ket{\wek{a}_{\wek{A}}}\Big)\Big(\sum_{\wek{b}}c^\alpha_{\wek{b}_{\wek{A}}}\bra{\wek{b}_{\wek{A}}}U_{\wek{A}}\Big)\\
&=\sum_\alpha{q_\alpha}\Big(\sum_{\wek{a}}c^\alpha_f(\alpha)\delta_{\wek{a},f(\alpha)}U^\dagger_{\wek{A}}\ket{\wek{a}_{\wek{A}}}\Big)\Big(\sum_{\wek{b}}c^{\alpha *}_f(\alpha)\delta_{\wek{b},f(\alpha)}\bra{\wek{b}_{\wek{A}}}U_{\wek{A}}\Big)\\
&=\sum_\alpha{q_\alpha}U^\dagger_{\wek{A}}\ket{f(\alpha)_{\wek{A}}}\bra{f(\alpha)_{\wek{A}}}U_{\wek{A}}.
\end{split}
\]
As every $U^\dagger_{\wek{A}}\ket{f(\alpha)_{\wek{A}}}$ is part of one and the same orthogonal basis, we get the claim.
\end{proof}
%\hfill $\Box$

Thus, every non-strictly classical state, either separable or entangled, does lead to the production of entanglement in the $\wek{A}:\wek{A'}$ cut, whatever the local rotation $U_{\wek{A}}$. On the other hand, such production of entanglement can be excluded by a proper local rotation $U_\wek{A}$ in the case of a strictly classically correlated state.

\section{Negativity of quantumness} \label{app:Nega}

The negativity is an entanglement monotone, defined for a bipartite state $\rho_{A:B}$ as $\cN(\rho_{A:B})=(||{\rho}_{A:B}^{T_{A}}||_{1}-1)/2$, with $\|X\|_1=\Tr\sqrt{X^\dagger X}$ the trace norm, and ${\rho}_{A:B}^{T_{A}}$ the partially transposed state. Thanks to the maximally correlated form, even in the multipartite case it is easy to calculate the eigenvalues of $\tilde{\rho}_{\wek{A}:\wek{A'}}^{T_{\wek{A}}}$, which are given by $\rho^{\wek{\cB}}_{\wek{i},\wek{i}}$, for all $\wek{i}$, and by the coherences $\pm|\rho^{\wek{\cB}}_{\wek{i},\wek{j}}|$ for $\wek{i}>\wek{j}$ (understood lexicographic order). Thus,
$\mathcal{N}(\tilde{\rho}_{\wek{A}:\wek{A'}})=(||\tilde{\rho}_{\wek{A}:\wek{A}}^{T_{\wek{A}}}||_{1}-1)/2=\left(\sum_{\wek{i}\neq\wek{j}}|\rho^{\wek{\cB}}_{\wek{i},\wek{j}}|\right)/2$, and we obtain another (quantitative) proof that $\tilde{\rho}_{\wek{A}:\wek{A'}}$ is entangled for any rotation $U_\wek{A}$ if and only if ${\rho}_{\wek{A}}$ is not classical. Indeed, by definition a non-classical state has some non-vanishing coherence $\rho^{\wek{\cB}}_{\wek{i},\wek{j}}$ for some $\wek{i}\neq\wek{j}$, in any basis $\wek{\cB}$. For a pure state $\ket{\psi}_{\wek{A}}=\sum_\wek{i}\Psi^{\wek{\cB}}_\wek{i}\ket{\wek{\cB}(\wek{i})}$, with $\Psi^{\wek{\cB}}_\wek{i}=\braket{\wek{\cB}(\wek{i})}{\psi}$, one has $\rho^{\wek{\cB}}_{\wek{i},\wek{j}}=\Psi^{\wek{\cB}}_\wek{i}\Psi^{\wek{\cB}*}_\wek{j}$, so that $\|\tilde{\rho}_{\wek{A}:\wek{A'}}^{T_{\wek{A}}}\|_1=\left(\sum_\wek{i} |\Psi^{\wek{\cB}}_\wek{i}|\right)^2$. In the bipartite case $\ket{\psi}_{AB}$ one has $\sum_\wek{i} |\Psi^{\wek{\cB}}_\wek{i}|=\sum_{i_1i_2}|\Psi^\wek{\cB}_{i_1i_2}|=\|\Psi^\wek{\cB}\|_{\ell_1}\geq \|\Psi^\wek{\cB}\|_1=\sum_k\sqrt{\lambda^\psi_k}$, with $\Psi^\wek{\cB}$ the matrix of coefficients, $\|\cdot\|_{\ell_1}$ and $\|\cdot\|_1$ the $\ell_1$-norm and trace norm, respectively, and $\ket{\psi}=\sum_k\sqrt{\lambda^\psi_k}\ket{\alpha_k}_{A}\ket{\beta_k}_{B}$ the Schmidt decomposition of $\ket{\psi}$. Thus, the \emph{negativity of quantumness} $Q_{\mathcal{N}}(\psi_{AB})$, i.e., the minimum negativity
%potential
of $\tilde{\rho}_{AB:A'B'}$, is  exactly equal to the standard negativity of $\ket{\psi}_{AB}$.
%Note that for a pure bipartite case the relative entropy of quantumness reduces to the entropy of entanglement $-\sum_k \lambda^\psi_k \log \lambda^\psi_k$~\cite{Bravyi2003}.
One can further consider the exemplary mixture
%\beq
%\label{eq:pseudoisotropic}
$
\rho(\psi,p)=(1-p)\frac{\openone}{d^2}+p\proj{\psi},
$
%\eeq
with $\openone/d^2$ the maximally mixed state. A straightforward calculation, taking again into account the maximally correlated structure of $\tilde{\rho}_{AB:A'B'}$, leads to $Q_{\mathcal{N}}(\rho(\psi,p))=p\mathcal{N}(\psi)$. So, as already observed in, e.g., \cite{groismanquantumness}, $\rho(\psi,p)$ is non-classical as long as $p>0$ and $\psi$ is entangled.

\section{Maximal non-classicality of separable states}
\label{app:maxsep}

Let us consider a separable state $\rho_{AB}=\rho^{\textup{sep}}_{AB}=\sum_mp_m\proj{\alpha^m}\otimes\proj{\beta^m}$, with $\{p_m\}$ a probability distribution, and $\ket{\alpha_m}$, $\ket{\beta_m}$ arbitrary pure states. Then,

%\subsection{Distillable entanglement}

%\begin{multline*}
%Q(\rho_{AB})\\
%\begin{split}
%&=\min_{\wek{\cB}}\Big(S(\rho_{AB}^{\wek{\cB}})-S(\rho_{AB})\Big)\\
%				&\leq \min_{\wek{\cB}}\Big(S(\rho_{AB}^{\wek{\cB}})-S(\rho_{A})\Big)\\
%				&=\min_{\wek{\cB}} \Big(S(\rho_{A}^{\cB_A}) + \sum_i \bra{\cB_A(i)}\rho_A\ket{\cB_A(i)} S(\sigma^{\cB_B}_{i})-S(\rho_{A})\Big)\\
%				&\leq\min_{\cB_B} \sum_i p_i^A S(\sigma^{\cB_B}_{i}),
%\end{split}
%\end{multline*}
\begin{eqnarray*}
Q(\rho_{AB})&=&\min_{\wek{\cB}}\Big(S(\rho_{AB}^{\wek{\cB}})-S(\rho_{AB})\Big)\\
				&\leq& \min_{\wek{\cB}}\Big(S(\rho_{AB}^{\wek{\cB}})-S(\rho_{A})\Big)\\
				&=&\min_{\wek{\cB}} \Big(S(\rho_{A}^{\cB_A}) + \sum_i \bra{\cB_A(i)}\rho_A\ket{\cB_A(i)} S(\sigma^{\cB_B}_{i})-S(\rho_{A})\Big)\\
				&\leq&\min_{\cB_B} \sum_i p_i^A S(\sigma^{\cB_B}_{i}),
\end{eqnarray*}
where
\[
\begin{split}
\sigma_i^{\cB_B}=\sum_j \frac{\bra{\cB_A(i)\cB_B(j)}\rho_{AB} \ket{\cB_A(i)\cB_B(j)}}{\bra{\cB_A(i)}\rho_A\ket{\cB_A(i)}}
\ket{\cB_B(j)}\bra{\cB_B(j)},
\end{split}
\]
and $\{p_i^A\}$ are the eigenvalues of $\rho_A$. The first inequality is due to the fact that for any separable state $S(\rho_{AB})\geq \max\{S(\rho_A),S(\rho_B)\}$~\cite{nielsenkempe}. The second inequality comes from choosing as particular basis $\cB_A$ an eigenbasis of $\rho_A$, so that $S(\rho_{A}^{\cB_A})=S(\rho_A)$. Now, this upper bound is equal to $\log_2d$ only if $\sigma_i^{\cB_B}$ is maximally mixed for all $i$, that implies that $\rho_B$ is also maximally mixed. Reversing the role of $A$ and $B$, we also find that $\rho_A$ must be maximally mixed for $Q(\rho_{AB})$ to be compatible with $\log_2d$. This means that the basis chosen in the second inequality is arbitrary, and we find that for the last line to be equal to $\log_2d$, it must be that $\bra{\cB_A(i)\cB_B(j)}\rho_{AB}\ket{\cB_A(i)\cB_B(j)}=1/d^2$ for all $\cB_A,\cB_B$ and all $i,j$. Thus it must be $\rho_{AB}=\openone/d^2$. But the latter state is classical. Thus, for any separable state that is not classical, we find that $Q(\rho_{AB})$ is less than $\log_2 d$, a value that is instead achieved by a maximally entangled state of $A$ and $B$.

\medskip
\section{Proofs of Theorems~\ref{prop:sep} and~\ref{prop:low-rank} }
\label{app:andreas1}

We will consider arbitrary local complete von Neumann measurements
\[
  M = \bigl( \proj{m_x} \bigr)_{x=1}^{d}, \quad
  N = \bigl( \proj{n_y} \bigr)_{y=1}^{d}
\]
on $A$ and $B$, respectively, and denote by $M$ also the completely positive trace-preserving projection
associated to the measurement:
\[
  M(\sigma) = \sum_x \proj{m_x} \sigma \proj{m_x},
\]
and likewise for $N$.  In the following let
$m = \lceil (\log_2d)^4 \rceil$.

\begin{beweis}[of Theorem~\ref{prop:sep}]
%\begin{proof}
The state $\sigma$ is almost identical to the information locking
states considered in~\cite{HLSW:rand}[Thm.~V.1, eq.~(64)],
except that there also $j$ is given in an extra register to $A$.
There it is shown that -- when $d$ is sufficiently large and with
high probability -- for any (projective) measurements on $A$ and $B$ with classical outputs $x$ and $y$,
respectively,
\beq
\label{eq:boundMI}
  I(x:y) \leq I(ij:y) \leq \text{const.}.
\eeq
This is because with our choice for $m$ ($n$ in ~\cite{HLSW:rand}), the parameter $\epsilon$ in the Eq. (66) of~\cite{HLSW:rand} can be chosen to scale as $1/\log_2 d$.
%high probability -- for any choice of bases $\cB_A$ and $\cB_B on $A$ and $B$, respectively,
%\[
%  I(A:B)_{\rho^{\cB_A\cB_B}} \leq I(ij:y) \leq \text{const.},
%\]
The bound \eqref{eq:boundMI} must hold true also for our state, since all we do is remove $j$
before the measurement.

Note however, that $x$ and $y$ have maximal entropy $\log_2d$, since
$\sigma_A = \sigma_B  =\1/d$ are both maximally mixed.
That means that
$S\bigl( \sigma_{AB}^\wek{\cB} \bigr) \geq 2\log_2d - \text{const.}$ as
claimed.

The upper bound on $S(\sigma_{AB})$ follows by observing that the rank of
$\sigma_{AB}$ can be at most $dm$.
\end{beweis}

%\section{Proof of Theorem~\ref{prop:low-rank}}
%\label{app:andreas2}

%We will consider arbitrary local complete von Neumann measurements
%\[
%  M = \bigl( \proj{m_x} \bigr)_{x=1}^{d}, \quad
%  N = \bigl( \proj{n_y} \bigr)_{y=1}^{d}
%\]
%on $A$ and $B$, respectively, and denote by $M$ also the completely positive trace-preserving projection
%associated to the measurement:
%\[
%  M(\sigma) = \sum_x \proj{m_x} \sigma \proj{m_x},
%\]
%and likewise for $N$.  In the following let
%$m = \lceil (\log_2d)^4 \rceil$.
%
%\begin{proof}

\begin{beweis}[of Theorem~\ref{prop:low-rank}]
The random state considered here is analysed in great detail already
in~\cite{HLW:aspects}, and we may refer to that paper for
technical results.

Since the rank of $\rho$ is bounded by $m$, the upper bound
on $S(\rho)$ is clear.

On the other hand, let us analyze the measure concentration
of the entropy $S\bigl( (M\ox N)\rho \bigr)$ -- first only for
a fixed pair $M$ and $N$. We use the elementary estimate
\[
\begin{split}
  S\bigl( (M\ox N)\rho \bigr) &\geq S_2\bigl( (M\ox N)\rho \bigr)\\
                              &=    -\log_2\sum_{x,y=1}^d \left( \Tr\psi(M_x \ox N_y \ox \1) \right)^2,
\end{split}
\]
where $S_2(\sigma)=-\log_2(\Tr \sigma^2)$ is the (quantum) Renyi entropy of order 2.
Hence, invoking the convexity of $-\log$ and the unitary invariance
of the distribution of $\psi$,
\[
%\begin{multline*}
\begin{split}
  \EE_\psi S\bigl( (M\ox N)\rho \bigr)
   &\geq -\log_2\EE_\psi \sum_{x,y=1}^d \left( \Tr\psi(M_x \ox N_y \ox \1) \right)^2 \\
       &=    -\log_2\left( d^2 \EE_\psi \bigl( \Tr\psi(\proj{0} \ox \proj{0} \ox \1) \bigr)^2 \right) \\
       &=    -\log_2\Big( d^2 \Tr\Big( \frac{\1+F_{ABC:A'B'C'}}{d^2m(d^2m+1)}\proj{00}_{AA'} \ox \proj{00}_{BB'} \ox \1_{CC'} \Big) \Big) \\
       &=     \log_2\frac{d^2m+1}{m+1}
        \geq  2\log_2d - \log\left(1+\frac{1}{m}\right).
\end{split}
\]
%\end{multline*}
The identity in the third line, $\Tr(F_{C:D}X_C\otimes Y_D))=\Tr(XY)$, where $F_{C:D}$ is the swap operator between $C$  and $D$ , is a standard trick.

The Lipschitz constant of the entropy $S\bigl( (M\ox N)\rho \bigr)$ can
be taken directly from the Appendix B of~\cite{HLW:aspects}: it is upper bounded by
$\sqrt{8}\log_2d^2$. Thus, by Levy's Lemma:
\beq
%\begin{multline}
  \Pr\left\{ S\bigl( (M\ox N)\rho \bigr)
                            < 2\log_2d - \log\left(1+\frac{1}{m}\right) - \epsilon \right\}\\
        \leq \exp\left( -c\frac{\epsilon^2}{32(\log_2d)^2} d^2m \right),
%  \label{eq:probability}
\eeq
for some constant $c>0$.

The rest of the proof is a net argument. If we consider $T$ possible basis pairs  $M_t$, $N_t$
($t=1,\ldots,T$), by the union bound we have:
\beq
%\begin{multline}
  \Pr\left\{ \exists t\ S\bigl( (M_t\ox N_t)\rho \bigr)
                            < 2\log_2d - \log\left(1+\frac{1}{m}\right) - \epsilon \right\}\\
        \leq T \exp\left( -c\frac{\epsilon^2}{32(\log_2d)^2} d^2m \right).
  \label{eq:probability}
\eeq
%\end{multline}
We shall show that it is enough to consider
\[
  T \leq \left( \frac{c'd^{3/2}(\log_2d)^2}{\epsilon^2} \right)^{4d^2}
\]
basis pairs, for which the probability in eq.~(\ref{eq:probability})
is $\ll 1$ for our choice of $m = \lceil (\log_2d)^4 \rceil$ and $d$ large
enough.
Each local measurement is decribed by an orthonormal
basis $(\ket{b_x})_{x=1}^d$. If we think of the vectors as column vectors
relative to some standard basis, we can arrange them in a $d\times d$
unitary matrix, the matrix rotating the standard basis to $(\ket{b_x})_{x=1}^d$.
Then, the claim is that on the unitary group $\mathcal{U}(d)$ with
operator norm distance, there exists a $\delta$-net of
\[
  T_0 \leq \left( \frac{c'd^{3/2}}{\delta} \right)^{2d^2}
\]
elements (see the net estimate below). Choosing $\delta = \epsilon^2/2(\log_2d)^2$ we obtain $T$
as $T_0^2$ (that is, using $T_0$ elements $\{M_t\}$ and $T_0$ elements $\{N_t\}$).

Now let $M$ and $N$ be arbitrary product
von Neumann projective measurements. We find $M_s$ and $N_t$ in the
net such that the unitaries of $M$ and $M_s$ are $\delta$-close,
and those of $N$ and $N_t$ likewise, which implies that
$M\ox N$ is $2\delta$-close to $M_s\ox N_t$ -- or in other words
the projection bases are related by a unitary $U$ that is
$2\delta$-close to the identity $\1$. But then it holds in trace distance $\frac{1}{2} \| U\rho U^\dagger - \rho \|_1 \leq 2\delta$ (see~\cite{HLSW:rand}[Lemma~II.4, Eq.~(17)]), and hence
%\begin{multline*}
\[
\begin{split}
  \left| S\bigl( (M\ox N)\rho \bigr) - S\bigl( (M_s\ox N_t)\rho \bigr) \right|
         &=    \left| S\bigl( (M\ox N)\rho \bigr) - S\bigl( (M\ox N)U\rho U^\dagger \bigr) \right| \\
       &\leq H_2(2\delta) + 2\delta \log_2d^2 \\
       &\leq (2\sqrt{2\delta}+2\delta) \log_2d^2
        \leq 4\epsilon.
\end{split}
%\end{multline*}
\]
where  $H_2$ denotes the binary entropy, and we have used the Fannes-Audenaert inequality [K. M. R. Audenaert, J. Phys. A: Math. Theor. \textbf{40}, 8127 (2007)] and the estimate $H_2(x) \leq 2\sqrt{x(1-x)}$.
Thus, we get directly
\[
%\begin{multline*}
  \Pr\left\{ \exists M \exists N\ S\bigl( (M\ox N)\rho \bigr)
                            < 2\log_2d - \log\left(1+\frac{1}{m}\right) - 5\epsilon \right\}
%   \begin{split}
           \leq \left( \frac{c'd^{3/2}(\log_2d)^2}{\epsilon^2} \right)^{4d^2}
                \exp\left( -c\frac{\epsilon^2}{32(\log_2d)^2} d^2m \right),
%\end{split}
%\end{multline*}
\]
and we're done.
\end{beweis}
%\end{proof}

\medskip

\begin{beweis}[of the net estimate]
From~\cite{HLW:aspects} we know that one can find an $\eta$-net on the
pure state vectors in $\CC^d$ (w.r.t.~the Euclidean norm) with
at most $\left(\frac{5}{\eta}\right)^{2d}$ elements. For each
vector $\ket{b_i}$ in the given basis, find an $\eta$-close
neighbour $\ket{b_i'}$. Of course this is not an orthogonal
basis in general, so we perform an orthogonalisation inspired by
the square-root measurement: define the operator
$B = \sum_{i=1}^d \proj{b_i'}$ and let
\[
  \ket{b_i''} = B^{-1/2} \ket{b_i'}.
\]
It is easy to see that if the $\ket{b_i'}$ are linearly independent, then
this defines an orthonormal basis. Linear independence is equivalent to
$B$ being invertible, which we also need for the above definition to make
sense.

Now observe $\| \proj{b_i'}-\proj{b_i} \| \leq 2\| \ket{b_i'}-\ket{b_i} \|_2 \leq 2\eta$,
hence
\[
  \| B- \1 \| =    \left\| \sum_i \left(\proj{b_i'}-\proj{b_i}\right) \right\|
              \leq \sum_i \| \proj{b_i'}-\proj{b_i} \| \leq 2d\eta,
\]
and consequently $\left\| B^{-1/2} - \1 \right\| \leq \frac{2d\eta}{1-2d\eta}$.

Putting all this together, and assuming $2d\eta \leq 1/2$,
we see that $\| \ket{b_i} - \ket{b_i''} \|_2 \leq (4d+1)\eta$.

An elementary estimate now shows that for the corresponding
unitary matrices $U$ and $U''$, $\| U-U'' \| \leq (4d+1)\sqrt{d}\eta$.
The number of different $U''$ encountered in this construction
is bounded by the $d$-th power of the net size on vectors we
started with, i.e.~$T_0 \leq \left(\frac{5}{\eta}\right)^{2d^2}$.

Letting $\eta = \frac{\delta}{c' d^{3/2}}$ concludes the proof.
\end{beweis}

%\begin{thebibliography}{9}

%\bibitem{bravyi}
%S. Bravyi, "Entanglement entropy of multipartite pure states", Phys. Rev. A
%67, 012313 (2003), arXiv:quant-ph/0205021

%\end{thebibliography}
\clearpage
\end{widetext}

\begin{thebibliography}{40}
\bibitem[{not()}]{notefaithful}
\bibinfo{note}{A measure of nonclassical correlations is faithful when it
  vanishes iff a state is strictly classically correlated.}
\end{thebibliography}
\end{document}